\documentclass[prl,twocolumn,floatfix,amsmath,nofootinbib,amssymb,floatfix]{revtex4}
\usepackage{graphicx,color,dcolumn,booktabs,bm}
\usepackage{longtable,lscape}
\usepackage{pdfpages}
\usepackage{txfonts}
\usepackage{overpic}
\usepackage{amssymb}
\usepackage{makecell}
\usepackage{indentfirst}
\usepackage{feynmf}   
\usepackage{slashed}  
\usepackage{cases}
\usepackage{color}
\usepackage{multirow}
\usepackage{threeparttable}
\usepackage{epstopdf}
\usepackage{enumerate}
\usepackage{graphicx,color,dcolumn,booktabs,bm}
\usepackage[colorlinks,
            citecolor=blue,
            anchorcolor=red,
            menucolor=red,
            linkcolor=red,
            filecolor=red,
            runcolor=red,
            urlcolor=blue,
            frenchlinks=red]{hyperref}

\begin{document}

\title{New type of hydrogenlike charm-pion or charm-kaon matter}
\author{Si-Qiang Luo$^{1,2,3,4,5}$}\email{luosq15@lzu.edu.cn}
\author{Zhan-Wei Liu$^{1,3,4,5}$}\email{liuzhanwei@lzu.edu.cn}
\author{Xiang Liu$^{1,3,4,5}$}\email{xiangliu@lzu.edu.cn}
\affiliation{
$^1$School of Physical Science and Technology, Lanzhou University, Lanzhou 730000, China\\
$^2$School of mathematics and statistics, Lanzhou University, Lanzhou 730000, China\\
$^3$Research Center for Hadron and CSR Physics, Lanzhou University and Institute of Modern Physics of CAS, Lanzhou 730000, China\\
$^4$Lanzhou Center for Theoretical Physics, Key Laboratory of Theoretical Physics of Gansu Province, Lanzhou University, Lanzhou 730000, China\\
$^5$Frontiers Science Center for Rare Isotopes, Lanzhou University, Lanzhou 730000, China}

\begin{abstract}
Borrowing the structures of the hydrogen atom, molecular ion, and diatomic molecule, we predict the nature of a new type of hydrogenlike charm-pion or charm-kaon matter that could be obtained by replacing the proton and electron in hydrogen matter with a charmed meson and a pion or a kaon, respectively. 
We find that the spectra of the atom, molecular ion, and diatomic molecule can be obtained simultaneously with the Coulomb potential for the hydrogen, the charm-pion, and the charm-kaon systems. The predicted charm-pion matter also allows us to explore the mass shift mediated by the strong interaction. For the charm-pion and charm-kaon systems, the strong interactions could lead to binding energy shifts. Our calculations suggests that the binding energy shifts in charm-pion systems are in the order of several to tens of eV. For the charm-kaon systems, the results are in the order of tens to hundreds of eV.
Exploring hydrogenlike charm-pion matter must lead to new demands for high-precision experiments.
\end{abstract}
\maketitle

\noindent{\it Introduction.---}As a representative of the precision frontier of particle physics, the exploration of exotic hadronic matter is always full of challenges and opportunities, which is as an effective approach to deepen our understanding of nonperturbative behavior of quantum chromodynamics (QCD). 
With the experimental progress in observing a series of charmoniumlike $XYZ$ states and several $P_c$ and $P_{cs}$ states in the last two decades \cite{Chen:2016qju,Liu:2013waa,Yuan:2018inv,Olsen:2017bmm,Guo:2017jvc,Hosaka:2016pey,Brambilla:2019esw,Chen:2022asf}, our knowledge of hadron spectroscopy has become rich. Especially recently, the LHCb, CMS and ATLAS collaborations continue to surprise us with the observation of $P_{\psi s}^\Lambda(4338)$ in $B^-\to J/\psi\Lambda \bar{p}$ \cite{LHCb:2022jad}, $T_{c\bar{s}0}^a(2900)^{0}$ and $T_{c\bar{s}0}^a(2900)^{++}$ in the $B^0\to \bar{D}^0D_s^+\pi^-$ and $B^+\to {D}^-D_s^+\pi^+$ weak decays \cite{LHCb:2022xob,LHCb:2022bkt}, and several enhancement structures in the di-$J/\psi$ invariant mass spectrum \cite{ATLAS:2022diJpsi,CMS:2022diJpsi}. Currently, we are entering new phase of construction of ``Particle Zoo 2.0". 

Since the birth of the quark model \cite{Gell-Mann:1964ewy,Zweig:1964ruk,Zweig:1964jf}, various types of hadronic configurations have been proposed, including multi-quark state, hadronic molecular state, hybrid and glueball. Among these exotic hadronic matters, the hadronic molecules, as loosely bound states composed of color singlet hadrons, have been extensively used to unravel the properties of these observed new hadronic states \cite{Chen:2016qju,Liu:2013waa,Yuan:2018inv,Olsen:2017bmm,Guo:2017jvc,Hosaka:2016pey,Brambilla:2019esw,Chen:2022asf}, where these  hadronic molecular states are formed by strong interaction. 
In fact, a positive charged hadron and a negative charged hadron can be attracted to each other via the Coulomb potential, which is similar to that of the atom. These hadronic matters are expected to have much smaller binding energies and larger sizes than the corresponding bound states induced directly by the strong interaction. Because of their 
special properties, the search for such bound states would be an interesting task.
{The typical examples are pion atoms, which are composed of a pion and a nucleus~\cite{Chatellard:1997nw,Sigg:1996qd,Thomas:1979xu}. In 2011, the SIDDHARTA Collaboration~\cite{SIDDHARTA:2011dsy,Bazzi:2012eq} observed a $N\bar{K}$ system. In addition, kaonic deuterium has been of great concern to both experimentalists and theorists~\cite{Curceanu:2013bxa,Zmeskal:2019ksw,Zmeskal:2015efj,Curceanu:2019uph,Liu:2020foc,Meissner:2006gx,Barrett:1999cw,Doleschall:2014wza}.}
Although great progress has been made on both the theoretical and experimental sides, we should introduce new insights to reveal new aspects of exotic hadronic matter as we stand at a new stage in the study of exotic hadronic matter.

\begin{figure}[htbp]
\centering
\includegraphics[width=8.6cm]{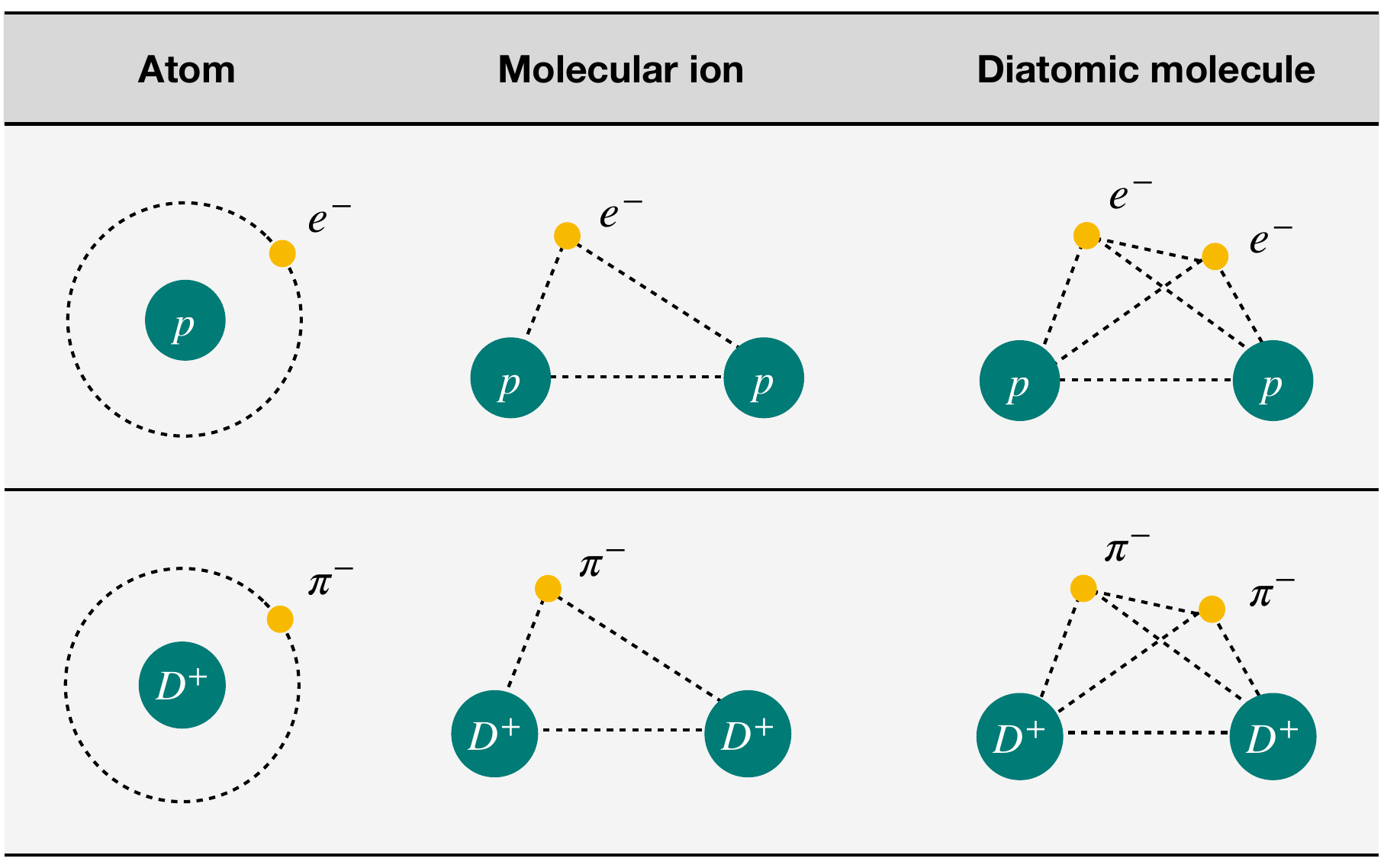}
\caption{The atom, molecular ion, and diatomic molecule in hydrogenous and charm-pion matters. Here, by replacing $\pi^-$ with $K^-$, we can obtain the corresponding charm-kaon matter.}
\label{fig:Handcharm}
\end{figure}

In this paper, we predict the existence of hydrogenlike charm-pion matters, where we can borrow some ideas by examining the hydrogen atom, the molecular ion, and the diatomic molecule, which are typical matter composed of the positive charged proton and the negative charged electron via the Coulomb potential, as shown in Fig. \ref{fig:Handcharm}. 
Taking the charmed meson $D^+$ and $\pi^-$ as the counterparts of $p$ and $e^-$, one gets a new kind of hydrogenlike charm-pion matter. As shown in Fig. \ref{fig:Handcharm}, there is a hydrogenlike atom ($H_{D^+}^{\pi^-}$), a molecular ion ($H_{D^+D^+}^{\pi^-}$), and a diatomic molecule ($H_{D^+D^+}^{\pi^-\pi^-}$) in charm-pion matter, which correspond to the known hydrogenous matter.
However, the proposed hydrogenlike charm-pion matter has some unique 
properties that are different from those of hydrogenous matter, 
which is the reason for attracting our attention. 
First of all, the wave functions of these hydrogenlike charm-pion matters are sufficiently  simple because the $D^+$ and $\pi^-$ are spin-0 particles, which can give the explicitly quantitative analysis of them. 
Second, {because of the Pauli principle, the wave functions of two protons or two electrons are antisymmetric. The situation is quite different in charm-pion matter, where the wave functions of two charmed mesons or two pions must be symmetric. In this paper, we perform a cross-calculation of the two systems to study how such symmetry affects the solution.}
In addition, for the charm-pion matter, the strong interaction as a perturbation contribution can lead to the energy shift of the discussed system. Thus, the predicted charm-pion 
matter can be used to study the nonperturbative behaviour of strong interaction, which could be tested by the experimental measurement. Of course, there is also charm-kaon matter as an extension of the focused hydrogenlike charm-pion matter, which will also be investigated in this paper.

\begin{table*}[htbp]
\caption{The binding energy $E$, root-mean-square radius $R$ of the atom, molecular ion, and diatomic molecule type systems. For the charm-pion and charm-kaon atoms, the decay widths are also given.}
\label{tab:DeltaEWidth}
\renewcommand\arraystretch{1.2}
\begin{tabular*}{172mm}{@{\extracolsep{\fill}}llcclclc}
\toprule[1.00pt]
\toprule[1.00pt]
&\multicolumn{3}{c}{Hydrogen}&\multicolumn{2}{c}{Charm-pion}                    &\multicolumn{2}{c}{Charm-kaon}                   \\
\Xcline{2-4}{0.75pt}
\Xcline{5-6}{0.75pt}
\Xcline{7-8}{0.75pt}
&                  &Expt.   &Theo.      &                                &Theo.              &                            &Theo.               \\
\midrule[0.75pt]
\multirow{4}{*}{Atom}
&$E$ (eV) &$-13.6$ &$-13.6$          &$E^{\rm QED}$ (keV)        &$-3.458$                &$E^{\rm QED}$ (keV)         &$-10.421$            \\
&$R$ (nm) &&0.09           &$R$ (fm)                   &360.6                 &$R$ (fm)                    &119.6              \\
&  &       &               &$\Delta E^{\rm OBE}$ (eV)  &$-4.4\sim-11.5$       &$\Delta E^{\rm OBE}$ (eV)   &$-44.7\sim-88.0$   \\
&   &      &               &$\Gamma$ (eV)              &0.03$\sim$0.47        &$\Gamma$ (eV)   &0.7$\sim$10.0      \\
\midrule[0.75pt]
\multirow{4}{*}{Molecular ion}
&$B$ (eV)          &$-16.25$ &$-16.20$    &$E^{\rm QED}$ (keV)             &$-3.848$            &$E^{\rm QED}$ (keV)         &$-11.182$            \\
&$R^{pp}$ (nm)     &       &0.11      &$R^{D^+D^+}$ (fm)               &613.0             &$R^{D^+D^+}$ (fm)           &259.2              \\
&$R^{pe^-}$ (nm)   &       &0.10      &$R^{D^+\pi^-}$ (fm)             &496.0             &$R^{D^+K^-}$ (fm)           &197.2              \\
&                  &       &          &$\Delta E^{\rm OBE}$ (eV)           &$-5.3\sim-14.6$   &$\Delta E^{\rm OBE}$ (eV)        &$-42.8\sim-87.2$   \\
\midrule[1.00pt]
\multirow{5}{*}{Diatomic molecule}
&$E$ (eV)          &$-31.65$ &$-31.60$     &$E^{\rm QED}$ (keV)             &$-7.517$  &$E^{\rm QED}$ (keV)           &$-21.889$      \\
&$R^{pp}$ (nm)     &      &0.127     &$R^{D^+D^+}$ (fm)               &574.3  &$R^{D^+D^+}$ (fm)             &187.3       \\
&$R^{e^-e^-}$ (nm) &      &0.076     &$R^{\pi^-\pi^-}$ (fm)           &435.4  &$R^{K^-K^-}$ (fm)             &214.3       \\
&$R^{pe^-}$ (nm)   &      &0.094     &$R^{D^+\pi^-}$ (fm)             &433.8  &$R^{D^+K^-}$ (fm)             &164.7       \\
&                  &      &          &$\Delta E^{\rm OBE}$ (eV)           &$-13.1\sim-32.0$       &$\Delta E^{\rm OBE}$ (eV)         &$-103.9\sim-209.7 $           \\
\bottomrule[1.00pt]
\bottomrule[1.00pt]
\end{tabular*}
\end{table*}

\noindent{\it The hydrogenlike charm-pion/kaon matter produced by the Coulomb interaction.---}In the charm-pion and charm-kaon systems, the electromagnetic interaction is well established, which is a simple Coulomb potential. For the systems of atom-type, the Schr\"odinger equation could be solved analytically. Explicitly, the binding energy and the root-mean-square radius of the ground state are
\begin{equation}\label{eq:solutiontwobody}
E=-\frac{1}{2}\mu e^4\alpha^2,~~~
R=\frac{\sqrt{3}}{\mu e^2\alpha}.
\end{equation}
In Eq.~(\ref{eq:solutiontwobody}), the $\mu$, $\alpha$, and $e$ are the reduced mass, the fine structure constant, and the charge, respectively. In addition, the wave functions in coordination and momentum spaces are given by
\begin{equation}
\psi({\bf r})=\frac{2(\mu e^2 \alpha)^{3/2}}{\sqrt{4\pi}}{\rm e}^{-\mu e^2 \alpha r},~~~
\psi({\bf p})=\frac{2\sqrt{2}}{\pi}\frac{(\mu e^2\alpha)^{5/2}}{((\mu e^2\alpha)^2+p^2)^2}.
\end{equation}

With the above preparations, one could easily obtain the binding energies of $D^+\pi^-$ and $D^+K^-$ with $-3.456$ and $-10.421$ keV, respectively (see Table~\ref{tab:DeltaEWidth}). The root-mean-square radius of the $D^+\pi^-$ and $D^+K^-$ could also be obtained by Eq.~(\ref{eq:solutiontwobody}) with 360.6 fm and 119.6 fm, respectively.

\begin{figure}[htbp]
\centering
\includegraphics[width=8.7cm]{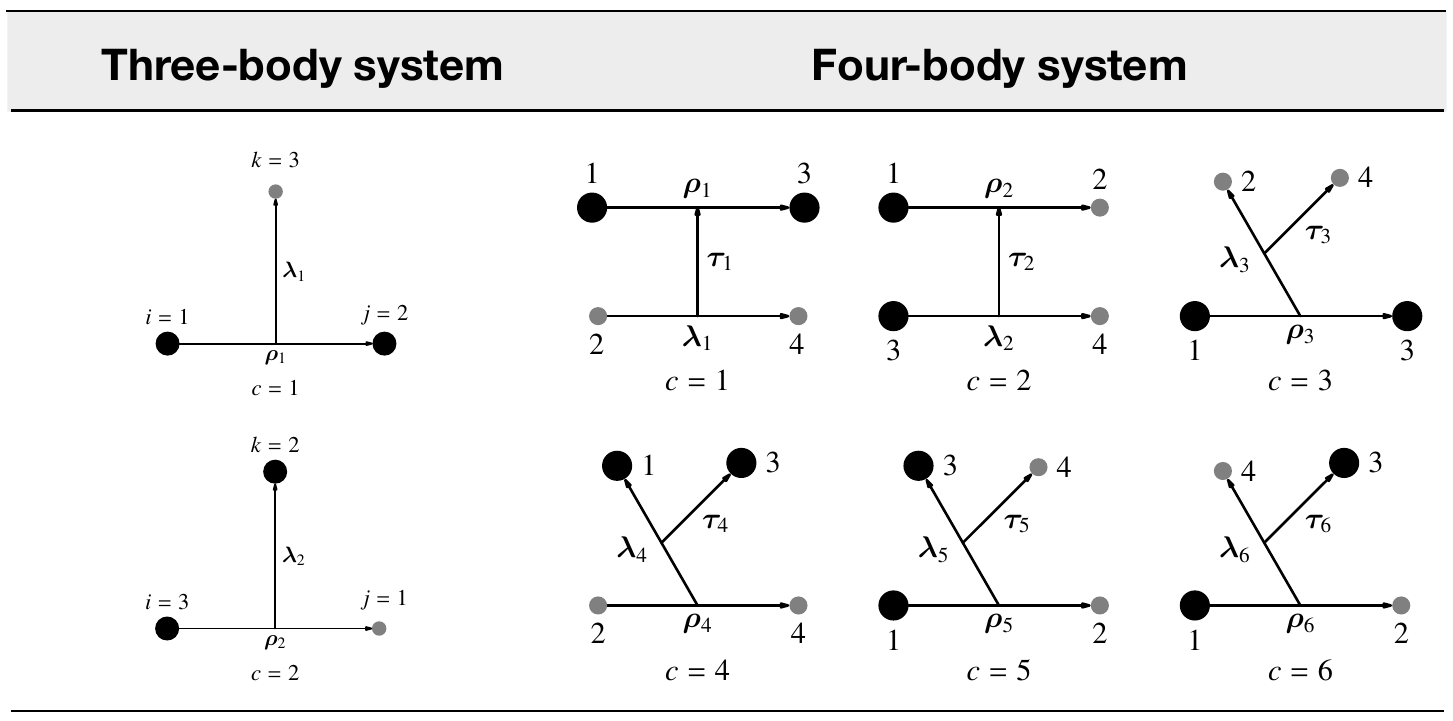}
\caption{The Jacobi coordinates of the three-body and four-body systems.}
\label{fig:JacobiCoordinates}
\end{figure}
However, for the molecular ion and diatomic molecule systems, the calculations are much difficult. In this work, we use the Gaussian expansion method (GEM)~\cite{Hiyama:2003cu} to solve the three- and four-body Schr\"odinger equations for the molecular ion and diatomic molecule systems, respectively. The corresponding Jacobi coordinates of the three- and four-body systems are shown in Fig.~\ref{fig:JacobiCoordinates}. By considering the symmetries in the two systems, the number of the Jacobi coordinates of a molecular ion in Ref.~\cite{Hiyama:2003cu} is reduced from 3 to 2. The 18 Jacobi coordinates of a diatomic molecule in Ref.~\cite{Hiyama:2003cu} are reduced to 6 groups.

To make the calculations more reliable, we first use the hydrogen molecular ion (${\rm H}_2^+$) and the hydrogen diatomic molecule (${\rm H}_2$) to test the solutions of the three- and four-body Schr\"odinger equations, respectively. As shown in Table~\ref{tab:DeltaEWidth}, the experimental binding energies of ${\rm H}_2^+$ and ${\rm H}_2$ are $-16.25$ and $-31.65$ eV, respectively, which are very close to our calculated results of $-16.20$ and $-31.60$ eV, respectively. This gives us confidence for the following calculations. The calculated binding energies of $D^+D^+\pi^-$, $D^+D^+K^-$, $D^+D^+\pi^-\pi^-$, and $D^+D^+K^-K^-$ are $-3.848$, $-11.182$, $-7.517$, and $-21.889$ keV, respectively. The root-mean-square radii of the charm-pion molecular ion and the diatomic molecule are roughly in the range of $430\sim610$ fm. For charm-kaon systems, the results are about in the range of $160\sim250$ fm.

For two identical particles, the wave functions are symmetric for spin-$0$, 
but antisymmetrical for spin-$\frac{1}{2}$. This leads to different treatments for hydrogen and charm-pion matters. In order to obtain the details of the bound state solutions more quickly, we use the following scaling trick. The Schr\"odinger equation with electromagnetic interaction could be written as
\begin{equation}
\left[\sum\limits_i\frac{p_i^2}{2m_i}+\sum\limits_{i<j}e_ie_j\frac{\alpha}{r_{ij}}\right]\psi=E\psi.
\end{equation}
If we normalize the mass $m_i$ with $m_i/{m_{\rm min}}$ and remove the $\alpha$ in the potential, the Schr\"odinger equation could be simplified as
\begin{equation}\label{eq:norschroeq}
\left[\sum\limits_i\frac{p_i^{\prime2}}{2m_i/(m_{\rm min})}+\sum\limits_{i<j}e_ie_j\frac{1}{r_{ij}^\prime}\right]\psi^\prime=E_s^\prime\psi^\prime.
\end{equation}
In the scheme, the solution depends only on the mass ratios. Since the dimensions of the masses are omitted by dividing $m_{\rm min}$, the ${\bf p}_i^\prime$, ${\bf r}_{ij}^\prime$, $E_s^\prime$, and $\psi^\prime$ also could be treated as dimensionless terms. And the practical solution could be obtained by the relationships
\begin{equation}
\begin{split}
&E=m_{\rm min}\alpha^2E_s^\prime,\\
&\psi\left(\{{\bf r}_{12},{\bf r}_{13},\cdots\}\right)=(m_{\rm min}\alpha)^{\frac{3(N-1)}{2}}\psi^\prime\left(m_{\rm min}\alpha\times\{{\bf r}_{12}^\prime,{\bf r}_{13}^\prime,\cdots\}\right),\\
&\psi\left(\{{\bf p}_{1},{\bf p}_{2},\cdots\}\right)=(m_{\rm min}\alpha)^{-\frac{3(N-1)}{2}}\psi^\prime\left((m_{\rm min}\alpha)^{-1}\times\{{\bf p}_{1}^\prime,{\bf p}_{2}^\prime,\cdots\}\right),
\end{split}
\end{equation}
where $N$ is the number of the particles. In the following, we fix $m_{\rm min}$ and only change the mass ratios to get $E_s^\prime$.

\begin{figure}[htbp]
\centering
\includegraphics[width=8.7cm]{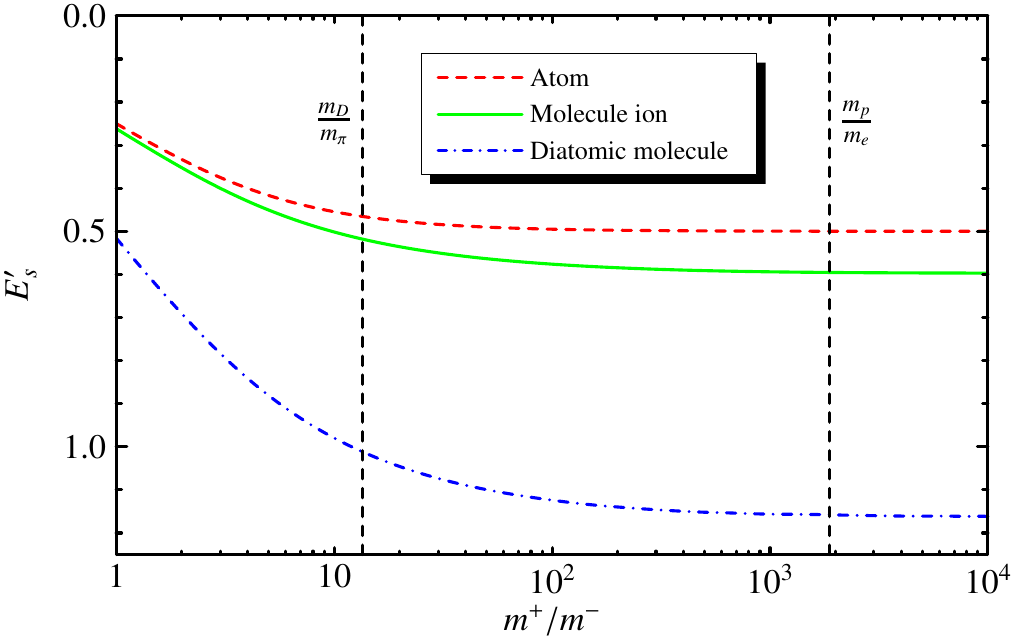}
\caption{The normalized binding energy $E_s^\prime$. Here, we define $m^-$,  i.e., the mass of the negative charged particle, as $m_{\rm min}$ in Eq.~(\ref{eq:norschroeq}).}
\label{fig:BE}
\end{figure}

The numerical results are shown in Fig.~\ref{fig:BE}. In the molecular and diatomic molecule systems, the symmetries of the hydrogen and charm-pion (kaon) the wave functions follow different rules and lead to different binding energies. However, this difference could be neglected. One possible reason is that the Coulomb potential is spin-independent. Thus whether spin-$\frac{1}{2}$ or -$0$ systems, one could refer the binding energy from Fig.~\ref{fig:BE}. Although it is difficult to find the differences between the two types of the systems, the symmetrization or antisymmetrization is still indispensable, otherwise we might get a nonphysical or singular solution.

\noindent{\it The mass shift mediated by the strong interaction.---}The mass shifts are important properties for a full understanding of the strong interactions. For example, we need the hadron-hadron potentials of the  strong interactions for $D^+$-$\pi^-$, $D^+$-$K^-$, $D^+$-$D^+$, $\pi^-$-$\pi^-$, and $K^-$-$K^-$. Here, we use one-boson-exchange (OBE) model to represent them. According to their symmetries, the strong interactions could occur through the exchange of $\sigma$ and $\rho$ mesons. For the $V^{D^+K^-}$, $V^{D^+D^-}$, and $V^{K^-K^-}$, we also consider the potential of $\omega$ exchange. With the effective Lagrangians in Refs.~\cite{Chen:2016ypj,Lu:2016nlp,Chen:2011cj}, we present the effective potentials explicitly, i.e.,
\begin{eqnarray}
V^{D^+\pi^-}&=&\frac{g_{DD\sigma}g_{\pi\pi\sigma}}{2m_\pi}Y_\sigma+\frac{1}{2}\beta g_V g_{\rho\pi\pi} Y_\rho,\label{eq:VOBE-1}\\
V^{D^+K^-}&=&\frac{g_{DD\sigma}g_{KK\sigma}}{2m_K}Y_\sigma+\frac{1}{2}\beta g_V g_{KK\rho} (Y_\rho-Y_\omega),\\
V^{D^+D^+}&=&-g_{DD\sigma}g_{DD\sigma}Y_\sigma+\frac{1}{4}\beta^2 g_V^2 (Y_\rho+Y_\omega),
\end{eqnarray}
\begin{eqnarray}
V^{\pi^-\pi^-}&=&-\frac{g_{\pi\pi\sigma}^2}{4m_\pi^2}Y_\sigma+g_{\pi\pi\rho}^2 Y_\rho,\\V^{K^-K^-}&=&-\frac{g_{KK\sigma}^4}{4m_K^2}Y_\sigma+g_{KK\rho}^2 (Y_\rho+Y_\omega)\label{eq:VOBE-2}
\end{eqnarray}
with
\begin{equation}\label{eq:Yi}
Y_i=\frac{{\rm e}^{-m_ir}}{4\pi r}-\frac{{\rm e}^{-\Lambda r}}{4\pi r}-\frac{\Lambda^2{\rm e}^{-\Lambda r}}{8\pi\Lambda}+\frac{m_i^2{\rm e}^{-\Lambda r}}{8\pi\Lambda}.
\end{equation}
In Eq.~(\ref{eq:Yi}), the $m_i$ is the mass of the exchanged light flavor boson. The $\Lambda$ is the cutoff parameter, defined in a monopole form factor
${\cal F}(q^2,m_i^2)=({\Lambda^2-m_i^2})/({\Lambda^2-q^2})$,
where the $q$ is the momentum of the exchanged boson. In previous work, the values of $\Lambda$ are around 1 GeV~\cite{Chen:2016ypj,Chen:2017jjn}. In this work, we extend the range of the $\Lambda$ with $1\sim2$ GeV to discuss how large the binding energy shifts are. In addition, we also need the coupling constants in Eqs.~(\ref{eq:VOBE-1})-(\ref{eq:VOBE-2}). The coupling constants for the interactions between the $D$ meson and the light flavor mesons are taken to be $\lambda=0.9$, $g_V=5.8$, and $g_{DD\sigma}=0.76$~\cite{Chen:2017jjn}. Using the central experimental widths of the $\rho$ and $\sigma$, we obtain $g_{\pi\pi\rho}=-6.01$ and $g_{\pi\pi\sigma}=-3.52$ GeV, where the signs are determined by the quark model. In addition, using the quark model relations, we have $g_{KK\rho}\approx g_{KK\omega}\approx\frac{1}{2}g_{\pi\pi\rho}$ and $g_{KK\sigma}\approx\frac{m_K}{2m_\pi}g_{\pi\pi\sigma}$.

\begin{figure}[htbp]
\centering
\includegraphics[width=8.6cm]{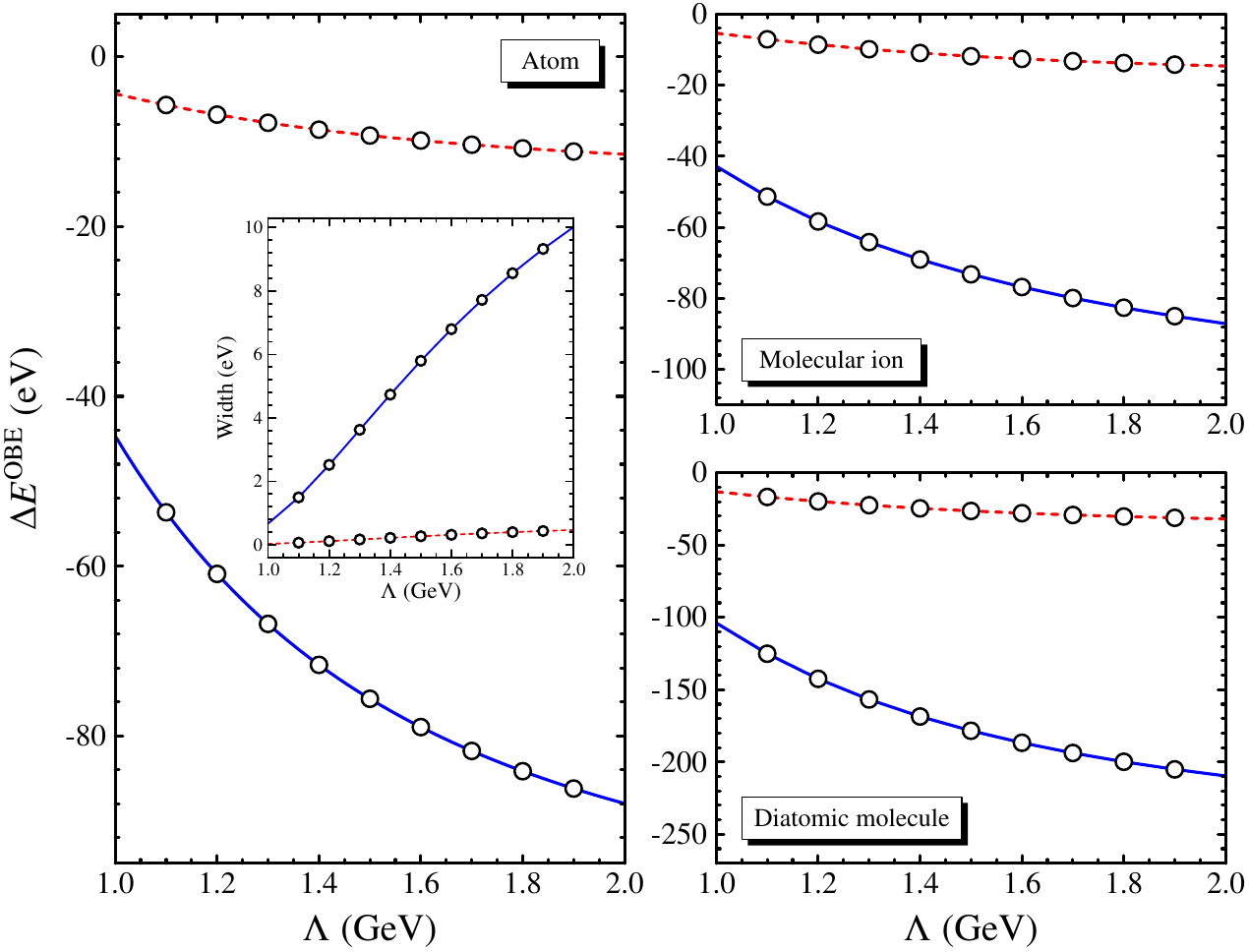}
\caption{The energy shift of the atom (left), the molecular ion (top right), and the diatomic molecule (bottom right) in the charm-pion and charm-kaon systems. The dashed dotted and solid dotted lines correspond to the results 
for the charm-pion and charm-kaon systems. For the charm-pion and charm-kaon atoms, we also show their decay widths.}
\label{fig:DeltaEWidth}
\end{figure}

We first study the atomic type of charm-pion and charm-kaon systems. The numerical results are presented in Table~\ref{tab:DeltaEWidth}. The details of the binding energy shifts as a function of the cutoff $\Lambda$ are shown in Fig.~\ref{fig:DeltaEWidth}. The energy shift of the $D^+\pi^-$ system is in the range of $-4.4\sim-11.5$ eV, which is approximately the same as in the $p\pi^-$ system. For the $D^+K^-$ system, the energy shift is in the range $-44.7\sim-88.0$ eV, which is smaller than that of the $pK^-$ system~\cite{SIDDHARTA:2011dsy}, but still comparable.

On the other hand, the strong decay behavior is also an important property. The decay widths could be calculated using the following approach. If two particles $AB$ form a bound state and are written as $[AB]$, the decay amplitude could also be calculated by the scattering process with
\begin{equation}\label{eq:decayMABtoCD}
\begin{split}
{\cal M}_{[AB]\to CD}=\int &\frac{1}{(2\pi)^{3/2}}\frac{\sqrt{2m_{[AB]}}}{\sqrt{2m_A}\sqrt{2m_B}}\psi_{AB}({\bf p}^\prime)\\
&\times{\cal M}_{AB\to CD}({\bf p},{\bf p}^\prime){\rm d}^3{\bf p}^\prime,
\end{split}
\end{equation}
where ${\cal M}_{AB\to CD}({\bf p},{\bf p}^\prime)$ is the scattering amplitude. Then the decay width is
\begin{equation}\label{eq:decayGABtoCD}
\Gamma_{[AB]\to CD}=\frac{p}{32\pi^2m_{[AB]}^2}\int |{\cal M}_{[AB]\to CD}|^2 {\rm d}\Omega.
\end{equation}
In Eqs.~(\ref{eq:decayMABtoCD})-(\ref{eq:decayGABtoCD}), the ${\bf p}^\prime$ and ${\bf p}$ is the momentum of $A$ and $C$, respectively. $\psi_{AB}({\bf p}^\prime)$ is the wave function of the bound state $[AB]$ in momentum space. The $D^+\pi^-$ and $D^+K^-$ systems have sufficiently simple decay modes. The $D^0\pi^0$ and $D^0\bar{K}^0$ are the main strong decay channels for $D^+\pi^-$ and $D^+K^-$, respectively. So we need the amplitudes of $D^+\pi^- \to D^0\pi^0$ and $D^+K^- \to D^0\bar{K}^0$, i.e.,
\begin{equation}
\begin{split}
i{\cal M}_{D^+\pi^- \to D^0\pi^0}=&[i^3g_{\pi\pi\rho}(p_{\pi^-}^0+p_{\pi^0}^0)]\frac{-ig_{00}+i\frac{q_0q_0}{m_\rho^2}}{q^2-m_\rho^2}\\
&\times[-i\sqrt{2}\beta g_V m_D]{\cal F}^2(q^2,\Lambda^2,m_\rho^2),\\
i{\cal M}_{D^+K^- \to D^0\bar{K}^0}=&[-i^3\sqrt{2}g_{KK\rho}(p_{K^-}^0+p_{\bar{K}^0}^0)]\frac{-ig_{00}+i\frac{q_0q_0}{m_\rho^2}}{q^2-m_\rho^2}\\
&\times[-i\sqrt{2}\beta g_V m_D]{\cal F}^2(q^2,\Lambda^2,m_\rho^2).
\end{split}
\end{equation}

We find that the phase spaces are extremely small, with the initial masses
only a few MeV above the thresholds. Thus, the two states are expected to be very narrow. As shown in Table~\ref{fig:DeltaEWidth} and Fig.~\ref{fig:DeltaEWidth}, the widths of $D^+\pi^-$ and $D^+K^-$ are predicted to be of the order of $\sim$0.1 eV and $\sim$1 eV. Such widths are so narrow when we compare the results with the commonly strong decays. With the binding energy shifts and widths, we could calculate the scattering lengths with~\cite{Meissner:2004jr}
\begin{equation}
\begin{split}
a_{D^+\pi^-}=&(0.067\sim0.175)+(0.0002\sim0.0036)i~{\rm fm},\\
a_{D^+K^-}=&(0.074\sim0.145)+(0.0006\sim0.0083)i~{\rm fm}.
\end{split}
\end{equation}

For the energy shifts or widths, the numerical results of $D^+K^-$ are much larger than $D^+\pi^-$ using the same cutoff $\Lambda$. This can be explained qualitatively by the root-mean-square radii of the two systems. According to Eq.~(\ref{eq:solutiontwobody}), the root-mean-square radii are inversely proportional to the reduced mass. So we could get the root-mean-square radius relation with $R_{D^+K^-}<R_{D^+\pi^-}$. Since the strong interaction is the typical short-range force, the system with a small root-mean-square radius can have larger strong interactions. In this way, it is not difficult to understand the large numerical differences between charm-pion and charm-kaon matters as shown in Fig.~\ref{fig:DeltaEWidth} and Table~\ref{tab:DeltaEWidth}.

For the molecular ion and the diatomic molecule, the strong interactions also increase the binding energy shifts. In the $H_{D^+D^+}^{\pi^-}$ and $H_{D^+D^+}^{K^-}$ systems, the shifts are  $-5.3\sim-14.6$ and $-42.8\sim-87.2$ eV, respectively. Such results are similar to the atomic scheme. For $H_{D^+D^+}^{\pi^-\pi^-}$ and $H_{D^+D^+}^{K^-K^-}$, the obtained energy shifts are $-13.1\sim-32.0$ and $-103.9\sim-209.7$ eV, respectively, which are about as twice large as the results in the atomic or molecular ion situation.

\noindent{\it Summary.---}It is fascinating to explore new forms of matter, which never cease to surprises us. Ordinary atoms make up our visible world, but we need to explore other kinds of atoms to reveal the essence of nature more deeply. The charm-pion and charm-kaon atoms or ions have many special properties that the ordinary atoms do not have. The ratios of the constituent masses of such structures are very different from those of ordinary atoms, and the systems composed of bosons and fermions would behave differently because of the statistics. It is also very interesting to study the couplings of electromagnetic and strong interactions in the formation of this new type of matter.

Following the conventional hydrogen matter, we study charm-pion and charm-kaon states with configurations of atom, molecular ion, and diatomic molecule. We discuss the binding energies and present the strong decay widths. The energy shifts due to the strong interaction are in the range of several to tens of eV for charm-pion systems, and are tens to hundreds of eV for the charm-kaon systems. In this work, the effect of the strong interaction is small, but important for us to understand its behavior.

In recent years, a number of high energy physics collaborations have produced many exciting results in hadrons physics. Such observations greatly advance our understanding of hadron structures and strong interactions. Although we predict the existence of a new types of hydrogenlike charm-pion or charm-kaon matter, it is still a challenge to search for such exotic matter experimentally. However, we believe that the great experimental challenges would inspire the experimentalists to improve the precision and break the detection limit in the future.

\noindent{\it Acknowledgement}.---We would like to thank Zhi-Feng Sun, Fu-Lai Wang and Li-Sheng Geng for useful discussions. This work is  supported by the China National Funds for Distinguished Young Scientists under Grant No. 11825503, the National Key Research and Development Program of China under Contract No. 2020YFA0406400, the 111 Project under Grant No. B20063, the National Natural Science Foundation of China under Grant No. 12175091, 11965016, 12247101, and the Project for Top-notch Innovative Talents of Gansu Province.


\begin{thebibliography}{}

\bibitem{Chen:2016qju}
H.~X.~Chen, W.~Chen, X.~Liu and S.~L.~Zhu,
The hidden-charm pentaquark and tetraquark states,
Phys. Rept. \textbf{639}, 1-121 (2016).

\bibitem{Liu:2013waa}
X.~Liu,
An overview of $XYZ$ new particles,
Chin. Sci. Bull. \textbf{59}, 3815-3830 (2014).

\bibitem{Yuan:2018inv}
C.~Z.~Yuan,
The XYZ states revisited,
Int. J. Mod. Phys. A \textbf{33}, no.21, 1830018 (2018).

\bibitem{Olsen:2017bmm}
S.~L.~Olsen, T.~Skwarnicki and D.~Zieminska,
Nonstandard heavy mesons and baryons: Experimental evidence,
Rev. Mod. Phys. \textbf{90}, no.1, 015003 (2018).

\bibitem{Guo:2017jvc}
F.~K.~Guo, C.~Hanhart, U.~G.~Mei\ss{}ner, Q.~Wang, Q.~Zhao and B.~S.~Zou,
Hadronic molecules,
Rev. Mod. Phys. \textbf{90}, no.1, 015004 (2018).

\bibitem{Hosaka:2016pey}
A.~Hosaka, T.~Iijima, K.~Miyabayashi, Y.~Sakai and S.~Yasui,
Exotic hadrons with heavy flavors: $X$, $Y$, $Z$, and related states,
PTEP \textbf{2016}, no.6, 062C01 (2016).

\bibitem{Brambilla:2019esw}
N.~Brambilla, S.~Eidelman, C.~Hanhart, A.~Nefediev, C.~P.~Shen, C.~E.~Thomas, A.~Vairo and C.~Z.~Yuan,
The $XYZ$ states: experimental and theoretical status and perspectives,
Phys. Rept. \textbf{873}, 1-154 (2020).

\bibitem{Chen:2022asf}
H.~X.~Chen, W.~Chen, X.~Liu, Y.~R.~Liu and S.~L.~Zhu,
An updated review of the new hadron states,
Rept. Prog. Phys. \textbf{86}, no.2, 026201 (2023).

\bibitem{LHCb:2022jad}
 [LHCb], Observation of a $J/\psi\Lambda$ resonance consistent with a strange pentaquark candidate in $B^-\to J/\psi\Lambda\bar{p}$ decays,
[arXiv:2210.10346 [hep-ex]].

\bibitem{LHCb:2022xob}
 [LHCb],
First observation of a doubly charged tetraquark and its neutral partner,
[arXiv:2212.02716 [hep-ex]].

\bibitem{LHCb:2022bkt}
 [LHCb], Amplitude analysis of $B^0 \rightarrow \overline{D}^0 D_s^+ \pi^-$ and $B^+ \rightarrow D^- D_s^+ \pi^+$ decays,
[arXiv:2212.02717 [hep-ex]].

\bibitem{ATLAS:2022diJpsi}
[ATLAS], Observation of an excess of di-charmonium events in the four-muon final state with the ATLAS detector, \href{https://atlas.web.cern.ch/Atlas/GROUPS/PHYSICS/CONFNOTES/ATLAS-CONF-2022-040/}{ATLAS-CONF-2022-040}.

\bibitem{CMS:2022diJpsi}
[CMS], Observation of new structures in the $J/\psi J/\psi$ mass
spectrum in pp collisions at $\sqrt{s}$ = 13 TeV, \href{http://cms-results.web.cern.ch/cms-results/public-results/preliminary-results/BPH-21-003/}{CMS-PAS-BPH-21-003}.

\bibitem{Gell-Mann:1964ewy}
M.~Gell-Mann,
A Schematic Model of Baryons and Mesons,
Phys. Lett. \textbf{8} (1964), 214-215.

\bibitem{Zweig:1964ruk}
G.~Zweig,
An SU(3) model for strong interaction symmetry and its breaking. Version 1,
CERN-TH-401.

\bibitem{Zweig:1964jf}
G.~Zweig,
An SU(3) model for strong interaction symmetry and its breaking. Version 2,
CERN-TH-412.

\bibitem{Chatellard:1997nw}
D.~Chatellard, J.~P.~Egger, E.~Jeannet, A.~Badertscher, M.~Bogdan, P.~F.~A.~Goudsmit, M.~Janousch, H.~J.~Leisi, E.~Matsinos and H.~C.~Schroeder, \textit{et al.}
X-ray spectroscopy of the pionic deuterium atom,
Nucl. Phys. A \textbf{625}, 855-872 (1997).

\bibitem{Sigg:1996qd}
D.~Sigg, A.~Badertscher, M.~Bogdan, P.~F.~A.~Goudsmit, H.~J.~Leisi, H.~C.~Schroeder, Z.~G.~Zhao, D.~Chatellard, J.~P.~Egger and E.~Jeannet, \textit{et al.}
The strong interaction shift and width of the ground state of pionic hydrogen,
Nucl. Phys. A \textbf{609}, 269-309 (1996)
[erratum: Nucl. Phys. A \textbf{617}, 526-526 (1997)].

\bibitem{Thomas:1979xu}
A.~W.~Thomas and R.~H.~Landau,
Pion - Deuteron and Pion - Nucleus Scattering: A Review,
Phys. Rept. \textbf{58}, 121 (1980).


\bibitem{SIDDHARTA:2011dsy}
M.~Bazzi \textit{et al.} [SIDDHARTA],
A New Measurement of Kaonic Hydrogen X-rays,
Phys. Lett. B \textbf{704}, 113-117 (2011).

\bibitem{Bazzi:2012eq}
M.~Bazzi, G.~Beer, L.~Bombelli, A.~M.~Bragadireanu, M.~Cargnelli, G.~Corradi, C.~Curceanu Petrascu, A.~d'Uffizi, C.~Fiorini and T.~Frizzi, \textit{et al.}
Kaonic hydrogen X-ray measurement in SIDDHARTA,
Nucl. Phys. A \textbf{881}, 88-97 (2012).

\bibitem{Curceanu:2013bxa}
C.~Curceanu, M.~Bazzi, G.~Beer, C.~Berucci, L.~Bombelli, A.~M.~Bragadireanu, M.~Cargnelli, A.~Clozza, A.~d'Uffizi and C.~Fiorini, \textit{et al.}
Unlocking the secrets of the kaon-nucleon/nuclei interactions at low-energies: The SIDDHARTA(-2) and the AMADEUS experiments at the DA$\Phi$FNE collider,
Nucl. Phys. A \textbf{914}, 251-259 (2013).

\bibitem{Zmeskal:2019ksw}
J.~Zmeskal, A.~Scordo, A.~Amirkhani, C.~Amsler, A.~Baniahmad, M.~Bazzi, G.~Bellotti, C.~Berucci, D.~Bosnar and M.~A.~Bragadireanu, \textit{et al.}
Probing Strong Interaction with SIDDHARTA-2,
JPS Conf. Proc. \textbf{26}, 023012 (2019).

\bibitem{Zmeskal:2015efj}
J.~Zmeskal, M.~Sato, S.~Ajimura, M.~Bazzi, G.~Beer, C.~Berucci, H.~Bhang, D.~Bosnar, M.~Bragadireanu and P.~Buehler, \textit{et al.}
Measurement of the strong interaction induced shift and width of the 1$s$ state of kaonic deuterium at J-PARC,
Acta Phys. Polon. B \textbf{46}, no.1, 101-112 (2015).

\bibitem{Curceanu:2019uph}
C.~Curceanu, C.~Guaraldo, M.~Iliescu, M.~Cargnelli, R.~Hayano, J.~Marton, J.~Zmeskal, T.~Ishiwatari, M.~Iwasaki and S.~Okada, \textit{et al.}
The modern era of light kaonic atom experiments,
Rev. Mod. Phys. \textbf{91}, no.2, 025006 (2019).

\bibitem{Liu:2020foc}
Z.~W.~Liu, J.~J.~Wu, D.~B.~Leinweber and A.~W.~Thomas,
Kaonic Hydrogen and Deuterium in Hamiltonian Effective Field Theory,
Phys. Lett. B \textbf{808}, 135652 (2020).

\bibitem{Meissner:2006gx}
U.~G.~Meissner, U.~Raha and A.~Rusetsky,
Kaon-nucleon scattering lengths from kaonic deuterium experiments,
Eur. Phys. J. C \textbf{47}, 473-480 (2006).

\bibitem{Barrett:1999cw}
R.~C.~Barrett and A.~Deloff,
Strong interaction effects in kaonic deuterium,
Phys. Rev. C \textbf{60}, 025201 (1999).

\bibitem{Doleschall:2014wza}
P.~Doleschall, J.~R\'evai and N.~V.~Shevchenko,
Three-body calculation of the 1s level shift in kaonic deuterium,
Phys. Lett. B \textbf{744}, 105-108 (2015).

\bibitem{Hiyama:2003cu}
E.~Hiyama, Y.~Kino and M.~Kamimura,
Gaussian expansion method for few-body systems,
Prog. Part. Nucl. Phys. \textbf{51}, 223-307 (2003).

\bibitem{Chen:2016ypj}
R.~Chen and X.~Liu,
Is the newly reported $X(5568)$ a $B\bar{K}$ molecular state?
Phys. Rev. D \textbf{94}, no.3, 034006 (2016).

\bibitem{Lu:2016nlp}
P.~L.~L\"u and J.~He,
Hadronic molecular states from the $K\bar{K}^{\ast}$ interaction,
Eur. Phys. J. A \textbf{52}, no.12, 359 (2016).

\bibitem{Chen:2011cj}
D.~Y.~Chen, X.~Liu and T.~Matsuki,
Two Charged Strangeonium-Like Structures Observable in the $Y(2175) \to \phi(1020)\pi^{+} \pi^{-}$ Process,
Eur. Phys. J. C \textbf{72}, 2008 (2012).

\bibitem{Chen:2017jjn}
R.~Chen, A.~Hosaka and X.~Liu,
Prediction of triple-charm molecular pentaquarks,
Phys. Rev. D \textbf{96}, no.11, 114030 (2017).

\bibitem{Meissner:2004jr}
U.~G.~Meissner, U.~Raha and A.~Rusetsky,
Spectrum and decays of kaonic hydrogen,
Eur. Phys. J. C \textbf{35}, 349-357 (2004).

\end{thebibliography}
\end{document}